\documentclass[10pt]{iopart}

\usepackage{cite}
\usepackage{graphicx}
\usepackage{appendix}

\begin{document}
\title[Simulating TWPAs using ADS]{Simulating the behaviour of travelling wave superconducting parametric amplifiers using a commercial circuit simulator}
\author{T. Sweetnam, D. Banys, V. Gilles, M.A. McCulloch, L. Piccirillo}
\address{Jodrell Bank Centre for Astrophysics, University of Manchester, UK}
\ead{thomas.sweetnam@manchester.ac.uk}
\vspace{10pt}
\begin{abstract}
Kinetic inductance travelling wave parametric amplifiers (KI-TWPAs) have been simulated using Keysight's Advanced Design System (ADS). A lumped element representation of a superconducting transmission line has been developed using nonlinear inductor elements to represent the current dependency of the intrinsic kinetic inductance. This provides a fast, simple and easily modifiable method for analysing the parametric processes that occur in nonlinear kinetic inductance devices such as parametric oscillators or amplifiers, while also allowing the accompanying RF network to be optimised. This methodology is used to model the behaviour of a select number of KI-TWPA designs operating in the 3-wave mixing (3WM) and 4-wave mixing (4WM) regimes. S-parameters and dispersion properties are extracted from large signal S-parameter simulations and the gain curve of each amplifier design is generated from harmonic balance simulations. Results of these are compared to the gain expression from the analytic solutions of the coupled mode equations, while making reference to the gain curves of the equivalent KI-TWPAs from literature.
\end{abstract}

%
%
%
%
\ioptwocol

\section{Introduction}
Following the demonstration of a dispersion engineered superconducting travelling wave parametric amplifier (TWPA) \cite{ho_eom_wideband_2012} research into these devices has accelerated, motivated by the large bandwidth, high dynamic range, and quantum noise level amplification that they have been shown to provide \cite{zmuidzinas_superconducting_2012,esposito_perspective_2021}. These characteristics make such paramps prime candidates for applications in quantum computing \cite{krantz_quantum_2019}, radio astronomy\cite{noroozian_superconducting_2018}, MKID readout\cite{bockstiegel_development_2014}, and the search for dark matter axions \cite{admx_collaboration_extended_2020,simanovskaia_symmetric_2020}. Parametric amplifiers exploiting the kinetic inductance nonlinearity of a superconducting film \cite{shu_nonlinearity_2021,malnou_three-wave_2021} or the nonlinearity of Josephson Junctions \cite{yaakobi_parametric_2013,zorin_josephson_2016} are active research areas, with amplifiers operating in the four wave mixing (4WM) \cite{chaudhuri_broadband_2017} and three wave mixing (3WM) \cite{vissers_low-noise_2016} regimes.\\
Currently there is no convenient, widely used, commercial toolset available for the design and simulation of TWPAs. Instead, the most frequently used method for simulating the performance of a TWPA involves solving a set of coupled mode equations for the structure being analysed. When only low order mixing products are included this method provides a reasonable estimate of the signal gain, but becomes increasingly complicated when the analysis is expanded to include higher order mixing products \cite{dixon_capturing_2019,erickson_theory_2017}. The coupled mode equation method requires that the dispersion properties of the structure are also known, which can be calculated analytically or by simulation. In addition, it is not trivial to take into account the transmission spectrum of the structure, or fabrication defects such as local hotspots or changes in transmission line geometry. Alternative methods such as solving the equations using a finite difference time domain model have also been used to model the behaviour of such an amplifier \cite{chaudhuri_simulation_2015,shan_parametric_2016}.\\
One circuit simulator that has been used to simulate TWPAs is WRspice\footnote{www.wrcad.com}, which includes a number of Josephson junction models. Studies of Josephson TWPAs (JTWPAs) have been carried out using this software in the exploration of energy flow from the pump to various generated harmonics \cite{dixon_capturing_2019} and the effect of junction fabrication tolerances on TWPA performance \cite{peatain_effect_2021}. The main limitation of WRspice is that the nonlinear components can only be simulated in transient mode, which can be computationally taxing to simulate and requires manual identification of mixing products. These software limitations can be mitigated by simulating the devices in harmonic balance mode in software packages such as Keysight ADS\footnote{www.keysight.com}, which is widely used in the design of active and passive RF circuits and has the advantage of being fast, user friendly and well documented. A similar method has recently been used to simulate parametrically-coupled networks in ADS \cite{naaman_synthesis_2021}.\\
This work will focus on the simulation of KI-TWPAs using their equivalent circuit models built in ADS to calculate the S-parameters, dispersion and gain of such devices without needing to set up and solve the high order coupled mode equations. This approach provides a convenient method of designing and simulating TWPAs with the ability to sweep design parameters and see their effect on amplifier performance\cite{banys_millimetre_2021}, whilst also predicting potential design flaws. The behaviour of the matching network can also be analysed, bringing TWPA design in-line with the industry standard tools that have been used in transistor-based low noise amplifier (LNA) research.\\
The mechanism behind parametric amplification in TWPAs will be outlined in Section \ref{sec:parametric_amplification}, and we will discuss how the gain can be calculated using the coupled mode equations. Section \ref{sec:simulation_setup} describes the simulation setup in ADS and the structure parameters needed to simulate signal gain of a TWPA, with results of these simulations for a variety of TWPA structures presented in Section \ref{sec:case_studies} and compared to the output from the coupled mode equations. The gain ripple seen in published devices is reproduced in Section \ref{sec:gain_ripple}, alongside an explanation for why it arises.\\

\section{Parametric amplification}
\label{sec:parametric_amplification}
A superconducting parametric amplifier generally consists of a transmission line such as coplanar waveguide or microstrip, fabricated from a superconducting thin film and using either the intrinsic kinetic inductance of the film or Josephson junctions to provide the source of nonlinearity required for amplification. This work focuses on devices that use the current-dependent nonlinear kinetic inductance. This arises due to the inertia of Cooper pairs within the superconductor, and is given by
\begin{equation}
    L_K(I) = L_{0}\left[1+ \frac{I^2}{I_*^2} + O(I^4)\right],
    \label{eqn:4wm_ki}
\end{equation}
where $I_*$ controls the scale of the nonlinearity and is of order of the critical current of the film \cite{semenov_effect_2020}. For a thin superconducting film of width $w < \lambda^2/t$ and thickness $t< \lambda_L$, which satisfies the nearly uniform current distribution assumption, an estimate of the scaling current is given by equating the kinetic and condensation energies:
\begin{equation}
    I_* \approx wt\sqrt{\frac{N_0\Delta^2}{\mu_0\lambda_L^2}},
    \label{eqn:Istar}
\end{equation}
where $N_0$ is the density of states at the Fermi level, $\lambda$ and $\lambda_L$ the perpendicular and London penetration depths \cite{shu_nonlinearity_2021}. For a more thorough description of $I_*$, Usadel theory is required \cite{anthore_density_2003}.\\
This nonlinearity allows for parametric signal gain to occur when the energy conservation condition (automatically satisfied via the generation of the idler) and the phase matching condition are met. The $I^2$ nonlinearity of the inductance permits 4WM gain to occur for a signal ($\omega_s$) in the presence of a strong pump ($\omega_p$) with the generated idler $\omega_i = 2\omega_p-\omega_s$. Addition of a DC bias $I_{DC}$ applied to the input of the amplifier modifies the inductance equation to become
\begin{equation}
    L_K(I) = L_0\left[ 1 + \frac{I^2_{DC}}{I^2_*} + 2\left( \frac{I_{DC}I_{RF}}{I^2_*} \right)    + \frac{I^2_{RF}}{I_*^2}        \right],
    \label{eqn:3wm_ki}
\end{equation}
which allows for 3WM to occur due to the term $\propto I_{RF}$, producing the idler $\omega_i = \omega_p - \omega_s$. Finding an expression for the gain of a KI-TWPA involves solving the wave equation for the current in a transmission line:
\begin{equation}
    \frac{\partial^2I}{\partial z^2} - \frac{\partial}{\partial t}\left[ L(I)C\frac{\partial I}{\partial t}  \right].
\end{equation}
Substituting for $L(I)$ using either Eq. \ref{eqn:4wm_ki} or \ref{eqn:3wm_ki} and solving the resulting expression with a travelling wave solution leads to a system of nonlinear equations, referred to as the coupled mode equations \cite{chaudhuri_simulation_2015}. 
In the simplest case which includes the signal$(s)$, pump$(p)$ and idler$(i)$ tones, an expression of the 4WM gain can be found by considering the phase mismatch between the three tones, $\Delta k = k_i + k_s - 2k_p $, which gives maximum gain when the phase matching condition is met:
\begin{equation}
    \Delta k + 2\Delta \phi = 0,
    \label{eqn:phase_matching}
\end{equation}
with $\Delta \phi = \frac{k_p}{8I_*^2}|I_{p0}|^2$, where $I_{p0}$ is the magnitude of the pump current and $k_{i,p,s}$ the wavenumber at each tone. When the condition in Eq. \ref{eqn:phase_matching} is achieved, the gain varies exponentially with the line length $l$,
\begin{equation}
G_s \approx 1 + \mathrm{sinh}^2(l\Delta \phi).
\label{eqn:matched_gain}
\end{equation}
In the 3WM case, the DC bias current modifies the condition for maximum gain, and it has been shown to be:
\begin{equation}
    k_p - k_s - k_i = -\frac{\epsilon I_{p0}^2}{8}(k_p - 2k_s -2k_i),
    \label{eqn:3wm_phase_matching}
\end{equation}
with $\epsilon = \frac{2I_{DC}}{I_*^2 +I_{DC}^2} $ \cite{malnou_three-wave_2021}.\\
To achieve maximum gain, the phase relationships in Eq. \ref{eqn:phase_matching} or \ref{eqn:3wm_phase_matching} between the pump, signal and idler must be maintained. This is done by modifying the dispersion of the transmission line, known as dispersion engineering. A planar TEM transmission line without any dispersion engineering features possesses a linear dispersion relation, $k=\frac{\omega}{v_{ph}}$, but this can be modified by periodically perturbing the line impedance or coupling resonators to the line.\\
To compensate for pump power induced phase slippage, early TWPAs introduced periodic perturbations in line impedance, producing stop bands in the transmission spectra of the line \cite{ho_eom_wideband_2012}. In the vicinity of the stop bands, the dispersion relation is modified. These designs placed the pump tone at an optimal point along this dispersion feature, restoring the phase matching condition. Phase matching can also be achieved using the dispersion features produced by coupling resonators to the transmission line.\\
\section{Simulation setup}
\label{sec:simulation_setup}
\begin{figure}
    \centering
    \includegraphics[width = 0.45\textwidth]{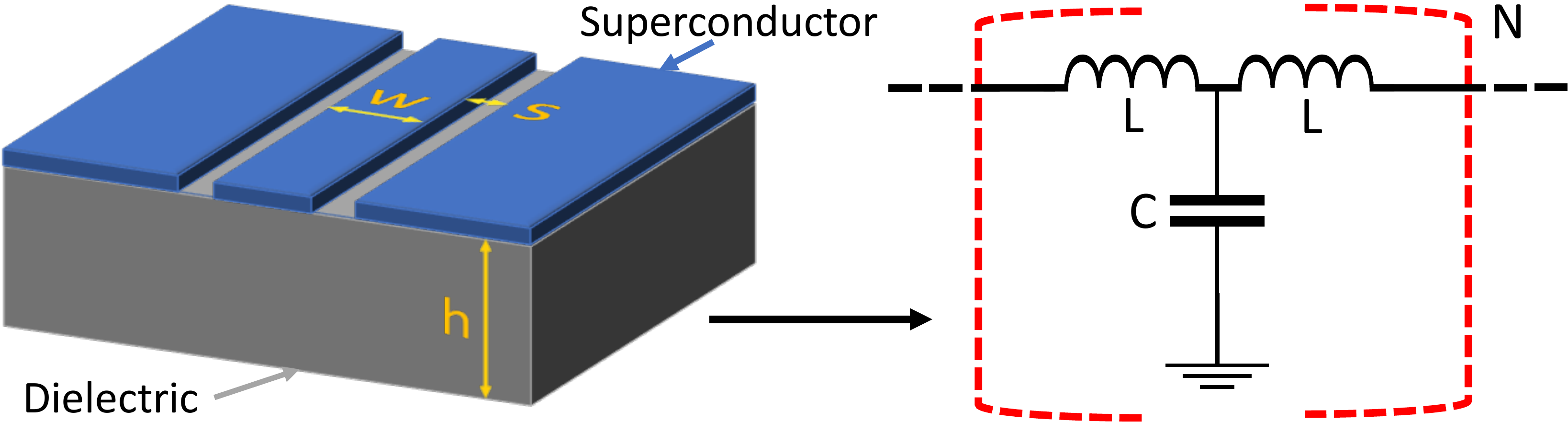}
    \caption{Coplanar waveguide represented by a lumped element transmission line model in a T network shape, with the inductors containing the contributions from both magnetic and kinetic inductance.}
    \label{fig:cpw}
\end{figure}
\subsection{Transmission line representation}
The behaviour of a superconducting parametric amplifier can be simulated in ADS by representing a nonlinear transmission line by its equivalent circuit model. For amplifiers using a coplanar waveguide (CPW) architecture, the geometry can be represented with lumped elements as shown in Figure \ref{fig:cpw}, with the capacitance and inductance per unit length calculated using a conformal mapping approach \cite{gao_physics_2008}. These elements are arranged in a T network, with each of the inductors having half of the inductance value calculated by the conformal mapping. This network arrangement allows us to include the forward and backward propagating waves in the simulations. More information about building the transmission line in ADS can be found in the Supp. Info.\\
The validity of the lumped element representation of a section of the line has been compared with the CPW components available in ADS, showing sufficiently similar behaviour as long as the line is split into enough sections to ensure that the lumped element cut off frequency,  $\omega_c = \frac{2}{\sqrt{LC}}$,  is above the amplifier's operating frequency. Splitting the LC line into smaller segments with $L_{segment} = L/N_{segments}$ and $C_{segment} = C/N_{segments}$ increases the cut off frequency by a a factor equal to the number of sections the initial line is split in to. This mitigates the reduction in transmission for frequencies close to $\omega_c$ but can introduce a numerical artefact in the simulations that results in a drop in gain at a point above the pump frequency. The cause of this effect is not currently well understood, but it is likely due to the method by which the harmonic balance simulation mode in ADS handles the lumped element representation of the transmission line at the step discontinuities in line width. Modelling this step in a more sophisticated way could potentially reduce this problem \cite{simons_modeling_1988}, but was not considered for this work. As a result, a number of segments that reduces the loss from the cut off frequency but also avoids this effect should be used. For simulations where the cutoff frequency was already much larger than the operating frequency of the amplifier, the line did not need splitting into segments.\\

\begin{figure}
    \centering
    \includegraphics[width = 0.5\textwidth]{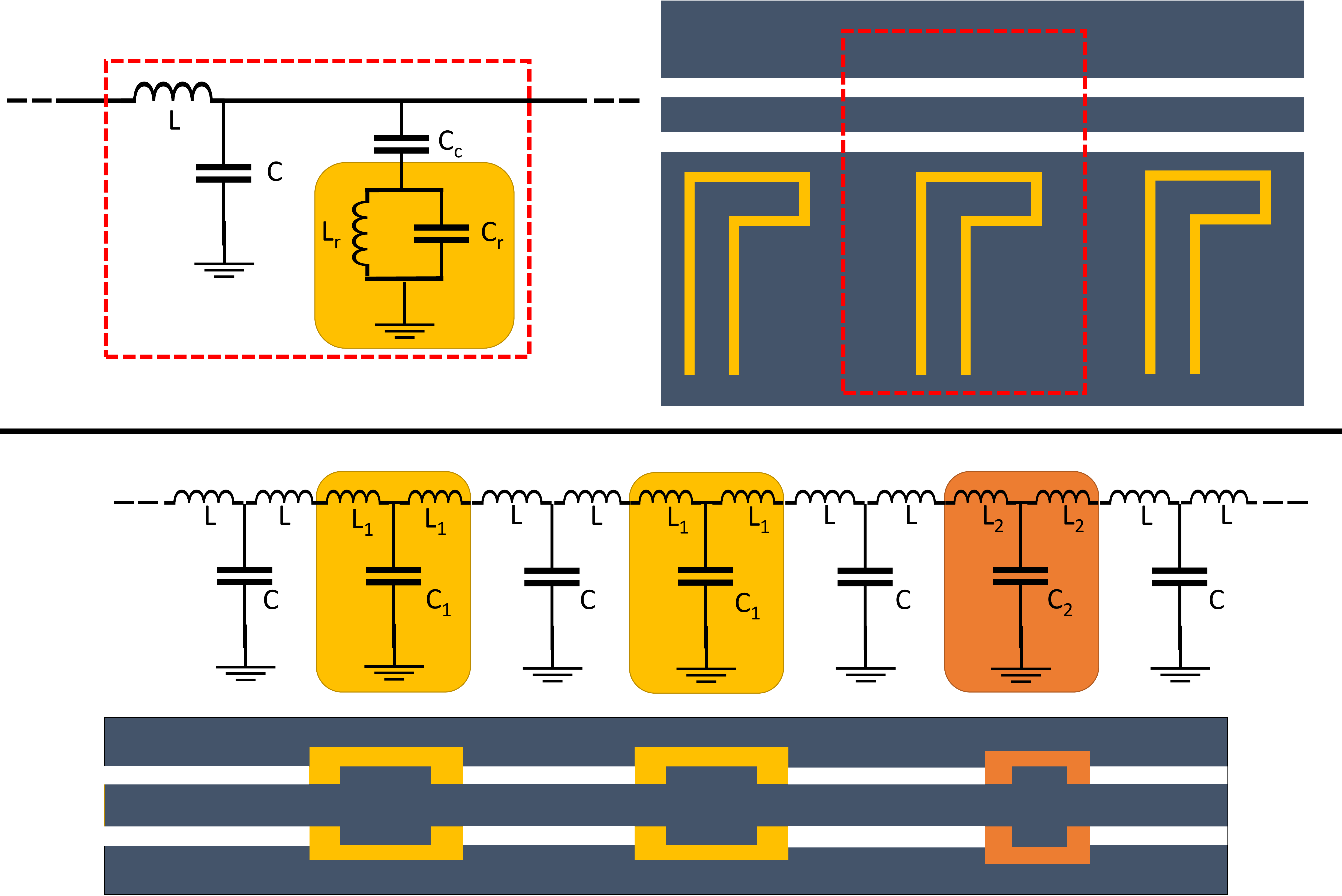}
    \caption{The lumped element representations of a unit cell of each of the TWPAs simulated. (above) A transmission line with resonators capacitively coupled to it as in \cite{obrien_resonant_2014}. (below) A periodically widened transmission line, with different widths corresponding to different $L$ and $C$ values as in \cite{ho_eom_wideband_2012} and \cite{vissers_low-noise_2016}.}
    \label{fig:tl_layout}
\end{figure}
\subsection{Superconductor properties}
\label{sec:super_properties}
The kinetic inductance of the superconducting film is included in the lumped element model of the transmission line by placing a current dependent nonlinear inductor in series with a simple inductor representing the magnetic inductance. This is implemented in ADS by using the non-linear inductor component that requires only the current-dependent non-linear inductance coefficients to be defined. Alternatively a more general component referred to as a symbolically defined device (SDD) can be used. The SDD requires for the time dependent voltage-current relationship to be defined. In the case of an inductor this is $V = \frac{d}{dt}(L(I)I)$. Both components produce the same result, with no noticeable difference in computing time. The expressions used for the nonlinear components in ADS can be found in the Supp. Info.\\
For simulations that more accurately represent measurements of the device in a laboratory environment, the properties of the superconducting film, kinetic inductance fraction $\alpha = \frac{L_K}{L_{tot}}$ and scaling current, must be found in addition to the $L$ and $C$ values of the transmission line that are extracted from the conformal mapping solutions. The zero current kinetic inductance fraction at temperatures ($T<<T_C$) can be calculated from the geometry of the coplanar waveguide using conformal mapping if the London penetration depth of the film is known. In the laboratory setting the position of a resonance or the stop bands compared to an equivalent non-superconducting device can be measured to find the shift due to kinetic inductance and therefore calculate $\alpha$. The scaling current, $I_*$, can be estimated from the properties of the film using Eq. \ref{eqn:Istar}, or measured in a nonlinear resonator by fitting the resonance frequency shift as a function of applied DC current \cite{shu_nonlinearity_2021}.\\
\subsection{Implementation in ADS}
The length of the line is modelled by defining a unit cell subcircuit consisting of the lumped elements in Figure \ref{fig:tl_layout}. Many of these unit cells are added in series to give the desired length of the line. This approach allows the repeating circuit parameters to be easily modified since the subcircuits can be nested within one another.
\begin{figure}
    \centering
    \includegraphics[width = 0.5\textwidth]{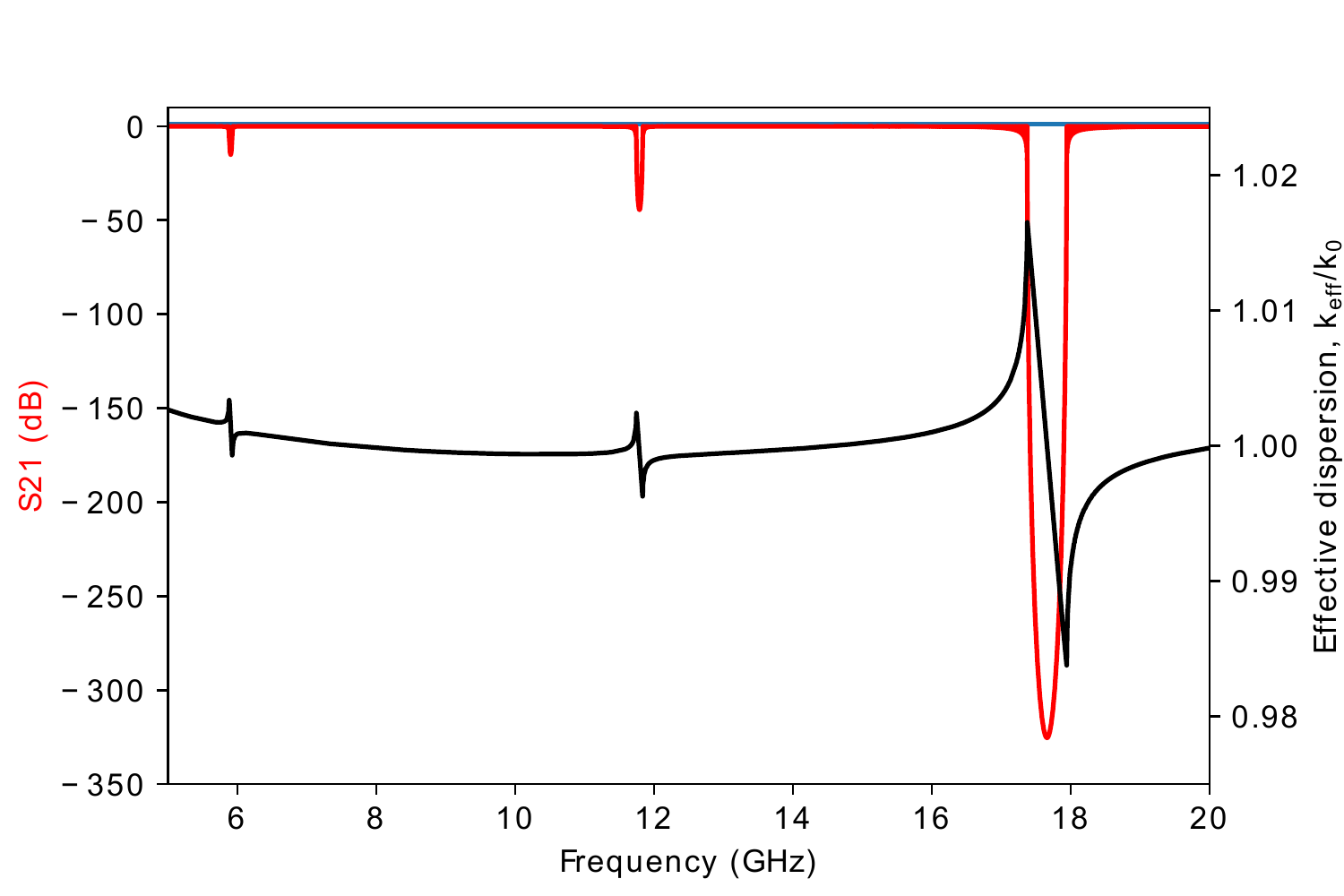}
    \caption{The deviation from linear dispersion, $k_0$, seen at the edges of the stop band, plotted with the stop bands. This data is from a simulation of the amplifier in \cite{ho_eom_wideband_2012}. Placing the pump along this dispersion feature allows phase matching to be maintained.}
    \label{fig:dispersion}
\end{figure}
Positions of the stop bands with a large pump tone were found using the large signal S parameter simulation (LSSP) mode, which provides the non-linear S-parameters using a harmonic balance back end. This method also allowed the dispersion properties of the structure to be analysed by plotting the unwrapped phase of the $S_{21}$ signal, shown in Figure \ref{fig:dispersion} for the amplifier simulated in section \ref{sec:eom}. The number of mixing products to be analysed within the circuit is controlled by the order set in the simulation parameters, allowing higher order harmonics to be included to track the flow of the pump power. Newton's method is used to solve for the Fourier coefficients of the solution, using the conservation of Kirchoff's law as a condition for the simulations to converge\footnote{Guide to Harmonic Balance in ADS - www.keysight.com}.\\
In the harmonic balance simulations, each mixing product within the simulation can be individually addressed at a specified point within the circuit, allowing for the input tone power flow to be tracked and the signal gain, generation of pump harmonics and other mixing products to be found. To find the gain of the amplifier, a probe point is set at the output port of the circuit and the amplitude of the mixing product at the frequency of the signal is found at that point. Subtracting the input signal power, which is defined at the input port, gives the signal gain.
\section{Case studies}
\label{sec:case_studies}
\subsection{4WM}
\label{sec:eom}
\begin{figure}
    \centering
    \includegraphics[width = 0.5\textwidth]{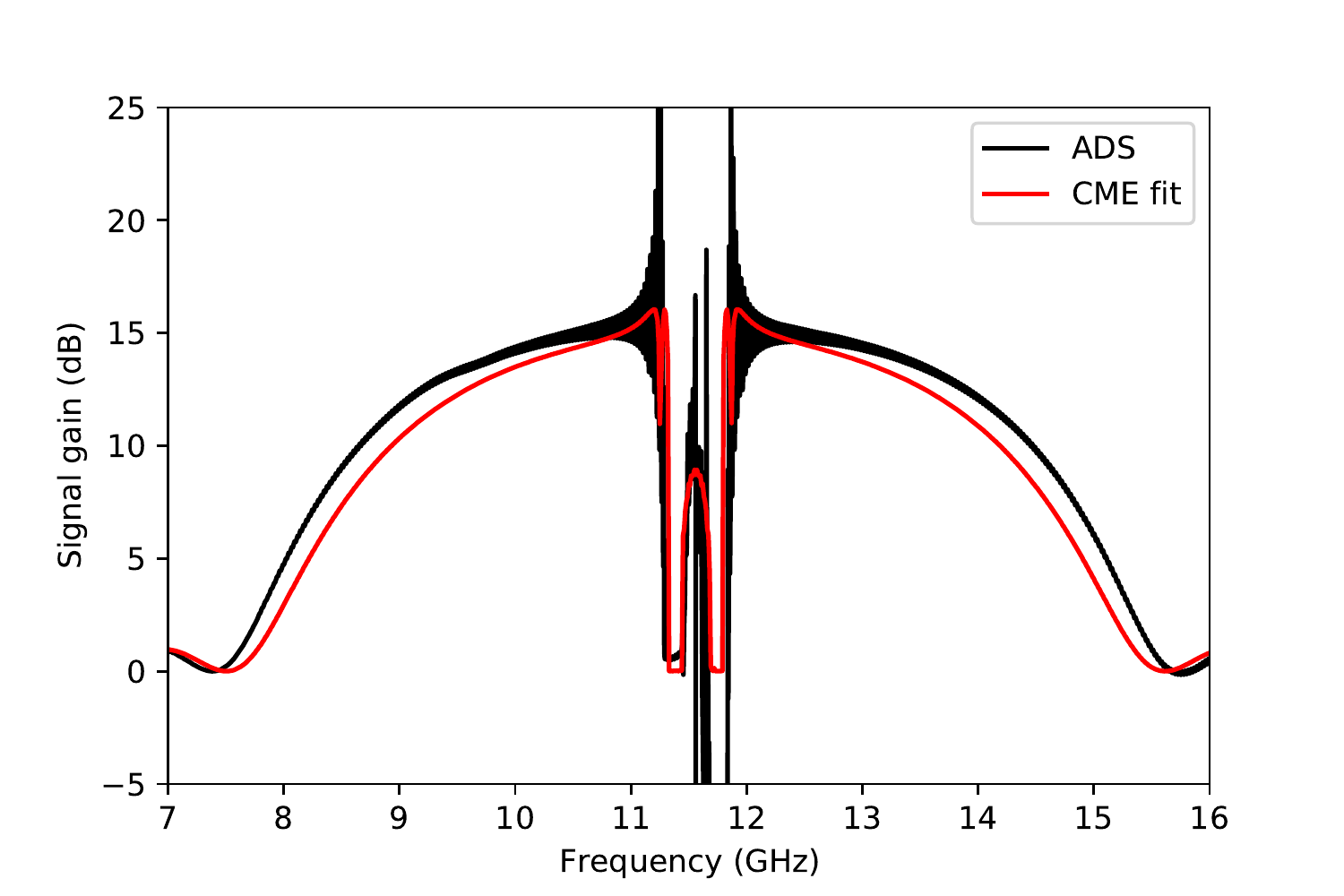}
    \caption{An ADS simulation of the KI-TWPA reported in \cite{ho_eom_wideband_2012}, compared to a coupled mode equations solution from \cite{chaudhuri_simulation_2015}. The CME fit is done using $I/I_*$ as the fitting parameter, with $I/I_* \approx 0.1$ in both the CME and ADS simulations.}
    \label{fig:eom_gain}
\end{figure}
The capability of ADS to accurately represent the behaviour of a KI-TWPA is tested by simulating the device reported by Ho Eom et al. \cite{ho_eom_wideband_2012}. These simulations are compared with the measurements in their paper and also with the results from the simulations in \cite{chaudhuri_simulation_2015}. The layout of this paramp consists of a 0.8 m long CPW transmission line interrupted by perturbations produced by widening the central conductor every 877 $\mathrm{\mu m}$. Every third perturbation is shorter, meaning that weaker anomalous dispersion features at 1/3 and 2/3 of the perturbation frequency are present. This was represented in ADS as shown in Figure \ref{fig:tl_layout}. The lumped element values were calculated via the method described in Section \ref{sec:super_properties} using the CPW dimensions that were provided in the paper, with each section of the line split into six segments to avoid the cutoff frequency problem. The zero current kinetic inductance fraction was inferred from the position of the stop band in the device at 11.8 GHz to be 93\%. The other important parameter governing the scale of the nonlinearity is $I_*$, which was estimated to be 8 mA using Eq. \ref{eqn:Istar}. The effect of including the quartic nonlinearity was also explored, but as the nonlinearity is dominated by the quadratic term this made no noticeable difference to the resulting gain.\\
\begin{figure}
    \centering
    \includegraphics[width = 0.5\textwidth]{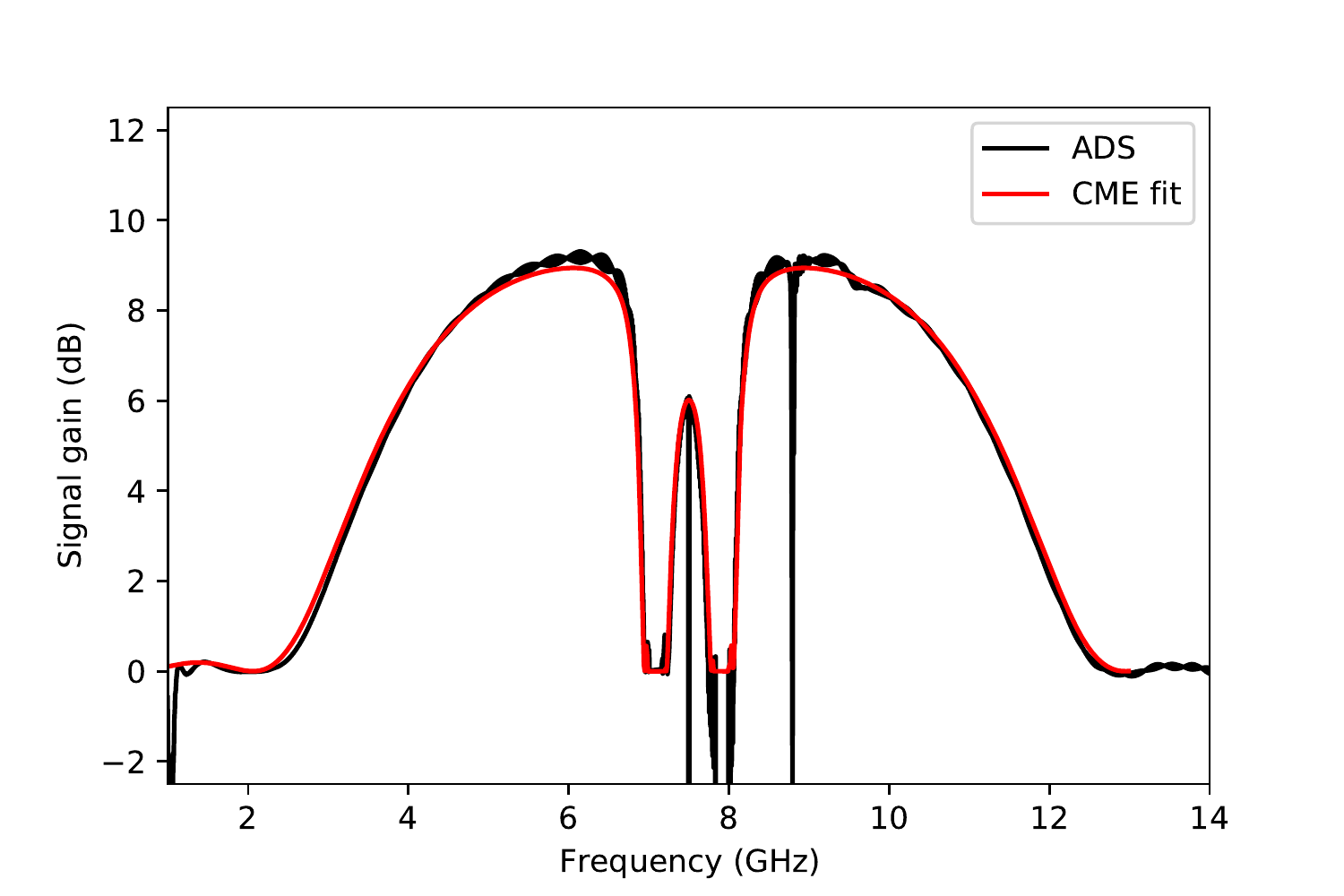}  
    \caption{Results from a 4WM ADS simulation of the KI-TWPA reported in \cite{vissers_low-noise_2016} using a periodically widened LC ladder model, compared to a coupled mode equations fit. The drop seen near 9 GHz is a numerical artefact arising from the effects due to cut off frequency described in section 2.1}
    \label{fig:vis_4wm}
\end{figure}
The transmission and effective dispersion of the line can be seen in Figure \ref{fig:dispersion}. The pump tone is placed on this dispersion feature, at 11.56 GHz with a power of -9.5 dBm, with the signal tone at -70 dBm to replicate the measurements in \cite{ho_eom_wideband_2012}. The power of the signal tone at the end of the transmission line is found and used to calculate the signal gain, shown in Figure \ref{fig:eom_gain} alongside a coupled mode equations fit using the dispersion data from Figure \ref{fig:dispersion}. The ADS simulation also shows strong agreement with the data reported in the paper, with gain above 10 dB extending over 2 GHz either side of the pump tone. Gain dropping to zero either side of the pump tone is characteristic of this type of amplifier, arising due to the signal being in the stop band above the pump, and the idler being in the stop band when the signal is below the pump, preventing any gain. The simulations have been able to reproduce the behaviour of the KI-TWPA without the need to solve the coupled mode equations, and include the gain ripple seen in the results from the original paper, even with the input ports perfectly matched to the line.\\
To match the levels of gain seen in the original device paper, the input power in this simulation is set to to -9.5 dBm compared to -9.4 dBm for their device. The small differences in magnitude and shape of the gain curve are likely due to our estimates of $I_*$ and the kinetic inductance fraction of the film, that are for pristine films and do not take into account fabrication uncertainties. In addition, the width of the perturbations in the ADS simulation and therefore the frequency width of the stop bands in the transmission of the device are estimated and may not be the same as the original device, leading to different behaviour in the vicinity of the stop band.\\
The coupled mode equations fit in Figure \ref{fig:eom_gain} is included using $I/I_*$ as the fitting parameter, with a value of 0.12, compared to $I/I_* = 0.09$ calculated for the ADS simulations. In \cite{dixon_capturing_2019} it was shown that adding higher harmonics to the coupled
mode equations results in power transfer to these, depleting the gain, but the much stronger pump tone used in our study has meant this effect is not seen even with the $8^{th}$ order mixing products used in the ADS simulations.
\subsection{3WM}
\begin{figure}
    \centering
    \includegraphics[width = 0.5\textwidth]{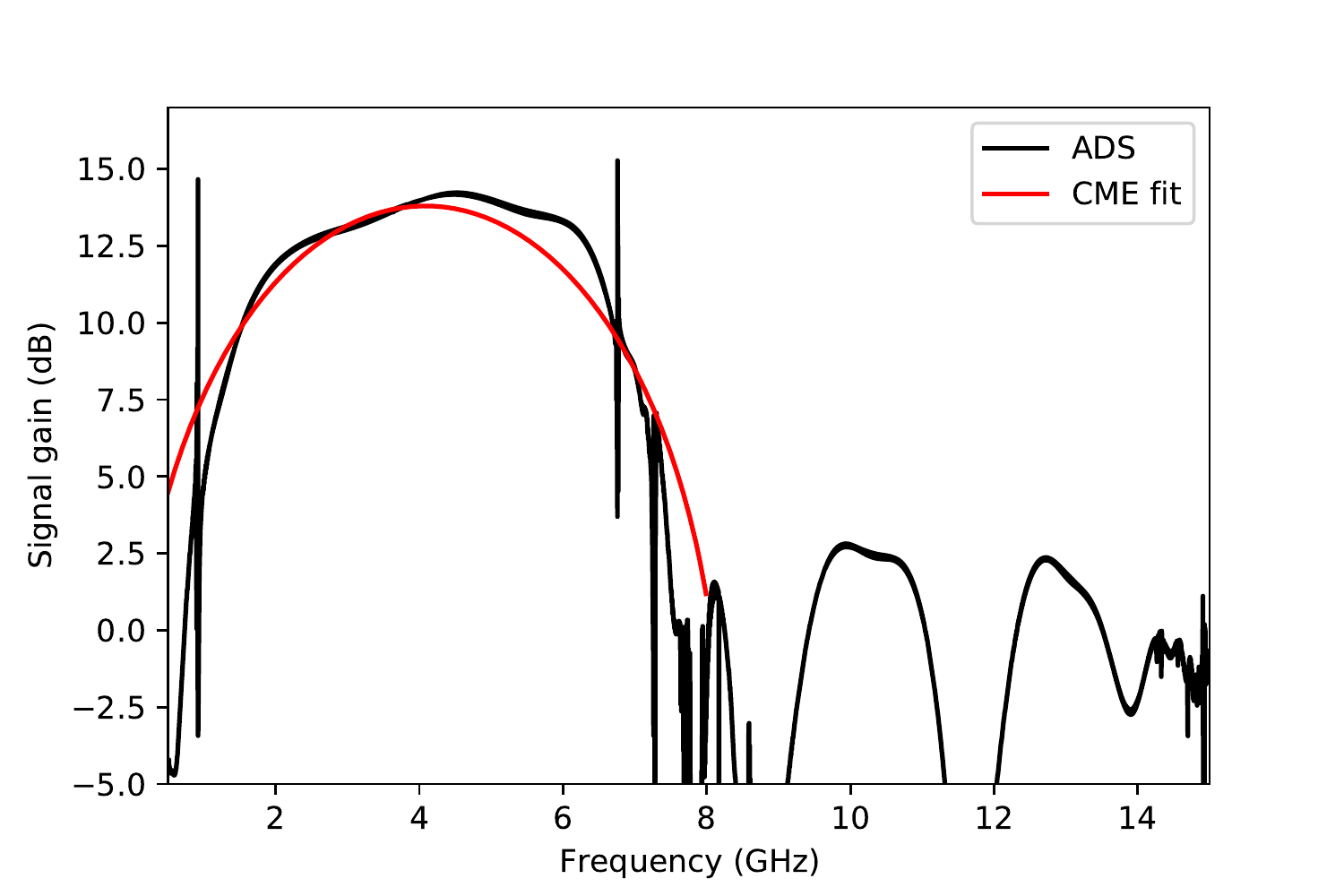}
    \caption{The ADS simulation results of the KI-TWPA described in \cite{vissers_low-noise_2016}, compared to a solution to the coupled mode equations in the 3WM regime considering only one idler \cite{malnou_three-wave_2021}.}
    \label{fig:vis_3wm}
\end{figure}
In addition to the 4WM processes described in the previous sections, 3WM amplifiers have also been simulated using this method. This involves modifying the kinetic inductance equation to include the DC current as in Eq. \ref{eqn:3wm_ki}, and adding a DC bias to the system. This bias was included in the model within the equation for kinetic inductance, rather than as a separate DC input component. The ADS model is designed to replicate the results of Vissers et al. \cite{vissers_low-noise_2016}, with periodic perturbations in the transmission line producing a stopband at 8 GHz. Initially, the gain was measured with the pump placed just below the stop band and no DC input; this produced the characteristic 4WM gain profile centred at the pump with a null around the pump frequency, shown in Figure \ref{fig:vis_4wm} for an input power of - 20 dBm and $I_* = 4$ mA.\\
A DC input of 0.6 mA was then applied to the circuit and the pump power reduced to -30 dBm. The presence of the DC bias required the pump to be moved to a higher frequency, above the stop-band, to achieve the new phase matching condition. This produced the 3WM gain plot centred at half the pump frequency shown in Figure \ref{fig:vis_3wm}. This generally follows the shape and gain values seen in the original device, including the region above the pump, with a much reduced ripple to that seen in their data.\\
\subsection{Resonant phase matching}
\begin{figure}
    \centering
    \includegraphics[width = 0.5\textwidth]{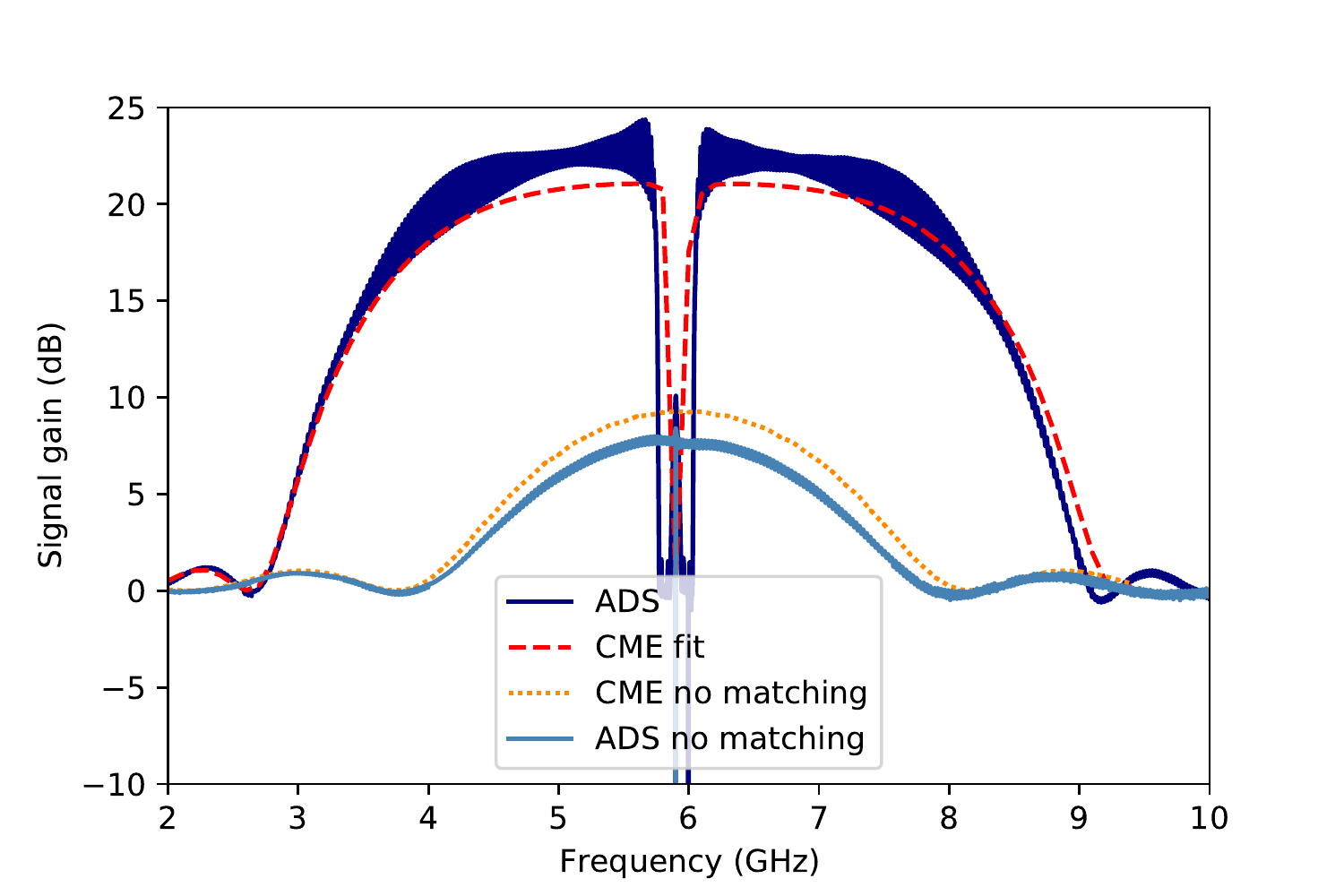}
    \caption{The gain produced by a long transmission line compared to the same transmission line with resonators periodically coupled to it, simulated using ADS and CME fits.}
    \label{fig:res_gain}
\end{figure}
Similarly to the method described in the previous sections, the gain response of a resonant phase matching amplifier was investigated to test if the simulation was able to describe the expected behaviour. Following a design described by O'Brien et al. \cite{obrien_resonant_2014}, a transmission line model with $\lambda/4$ resonators at 6 GHz periodically coupled to it through capacitive gaps was made, using kinetic inductance as the source of the nonlinearity rather than the Josephson junctions considered in their paper. The superposition of the resonant features produces a stopband at 6 GHz, which allows the phase matching condition to be met when the pump is placed at the optimal point along the corresponding dispersion feature.\\
The parameters of the lumped elements shown in Figure \ref{fig:tl_layout} are the same as in \cite{obrien_resonant_2014}, other than the kinetic inductance fraction which was set to 95 \%, with $I_*$ at 4 mA. A signal tone at -100 dBm was input and the pump power tuned to -17 dBm, which produces gain levels similar to those seen in the original paper. Removing the coupled resonators and repeating the simulation allowed the phase matched case to be compared to the gain produced in a nonlinear transmission line with no perturbations and therefore no stopband present. Shown in Figure \ref{fig:res_gain}, the benefits of phase matching are clearly seen, achieving much higher gain over a larger bandwidth for the same input power. The large gain ripples seen approaching the resonant frequency of the coupled resonators are likely due to an impedance mismatch; the ports are matched to the impedance of the line but the effect of the resonators is not considered, leading to the mismatch.\\

\section{Gain ripple}
\label{sec:gain_ripple}
\begin{figure}
    \centering
    \includegraphics[width = 0.5\textwidth]{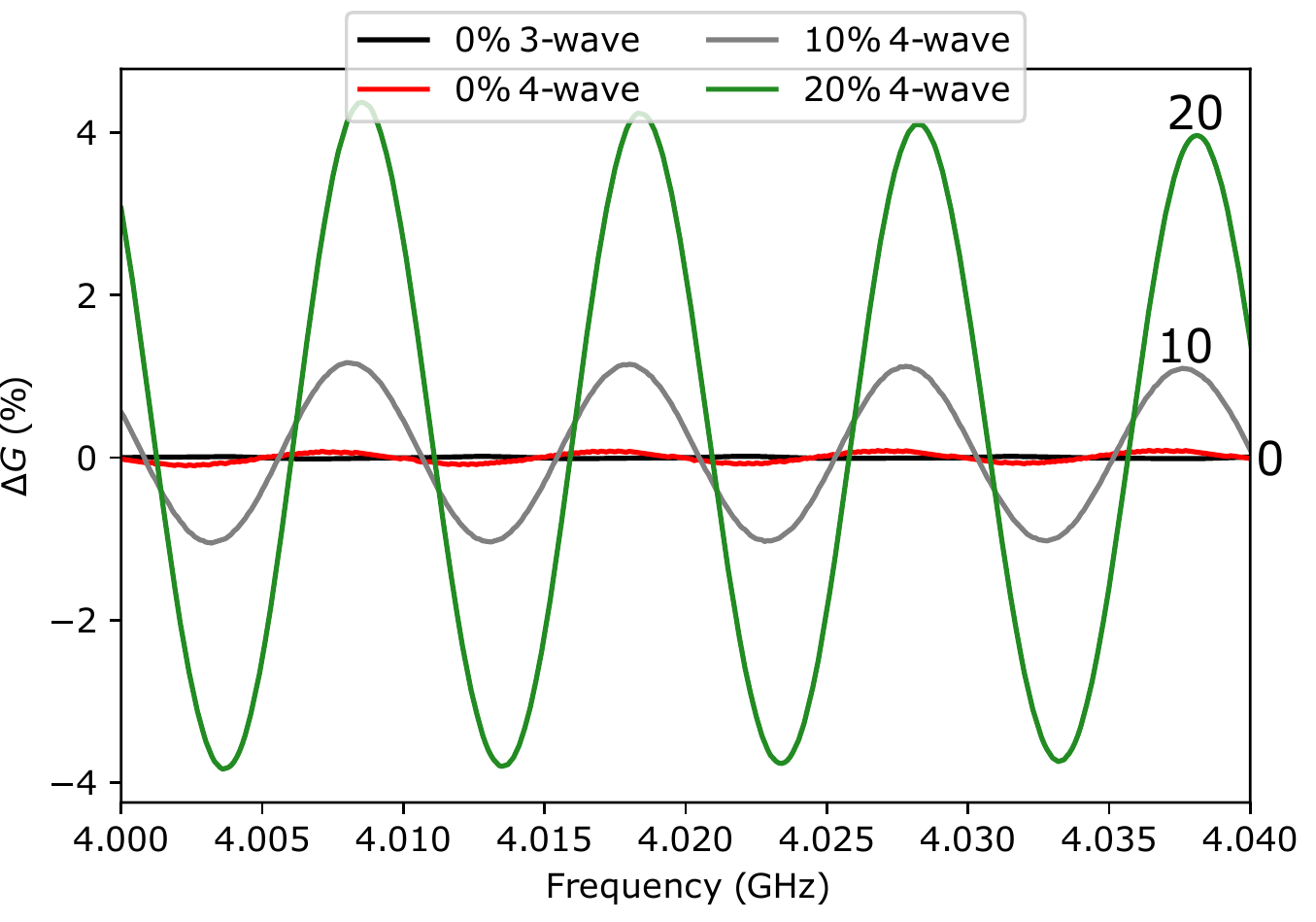}
    \caption{The gain ripple plotted as a percentage seen above 4 GHz for a range of impedance mismatches in the simulations from Figures \ref{fig:vis_4wm} and \ref{fig:vis_3wm}. The impedance mismatch increases the ripple as expected, with the 3WM regime showing the smallest ripple}
    \label{fig:ripple}
\end{figure}
A disadvantage of the parametric amplifiers seen in literature is the presence of a ripple in the gain spectrum, often a few dB in size. This is likely due to the impedance transformers used to match from the feed to the transmission line, or poor impedance matching elsewhere that causes reflected waves due to the mismatch and has been described in detail in \cite{kern_transition_2022}. The gain ripples seen in the ADS simulations are smaller than those seen in the works these amplifiers are based on, but are still present. The simulations in Figure \ref{fig:ripple} are performed with the input ports perfectly impedance matched to the line; with an impedance mismatch added by adjusting the port impedance the gain ripples increase in magnitude as expected. The ripples persisting in the perfectly matched line suggests that they are intrinsic to the structure rather than just from the mismatch.\\
Erickson and Pappas \cite{erickson_theory_2017} describe the origin of the intrinsic gain ripple in a 4WM amplifier: due to the stop band discontinuity just above the pump frequency, the behaviour of the dispersion frequency bands is different above and below the pump. As the signal is below the pump and the 4WM idler above the pump, they lie in different dispersion bands and the  relationship between phase velocity and frequency is different for both, meaning the conservation of momentum given by $k_s+k_i = 2k_p$ is not maintained as the signal frequency is varied, resulting in fluctuations in the gain amplitude. Their model predicts that as the stop band size is reduced the gain ripples decrease in size, due to the reduced differences between the dispersion bands.\\
Amplifiers operating in the 3WM regime tend to be less affected by these gain ripples. In this case, the idler and signal both lie in the same dispersion band below the pump frequency. Therefore, they show the same relationship between frequency and phase velocity and as a result are able to maintain the conservation of momentum condition, resulting in no ripple. The small gain ripple in the 3WM case seen in Figure \ref{fig:ripple} is likely due to 4WM processes that are occurring alongside the dominant 3WM processes, resulting in ripples that are smaller than in the purely 4WM case due to the smaller applied RF powers.\\

\section{Conclusions}
This work has demonstrated the capability of a lumped element based model to simulate the behaviour of various kinds of kinetic inductance parametric amplifier in a circuit simulator. Whether operating in a 4WM mode or 3WM mode with a DC bias, our simulations qualitatively reproduce the gain plots seen in existing devices to a reasonable degree of accuracy. Finding the gain of a device without needing to solve the coupled mode equations, with the ability to run parametric sweeps to optimise amplifier designs makes this a useful tool in the device design and fabrication process.\\
This paper focused on simulating amplifiers with a CPW architecture that use kinetic inductance as the source of the nonlinearity, but the method can be easily extended to simulate amplifiers constructed from different transmission lines or using alternative sources of nonlinearity such as Josephson junctions, by simply modifying the equation used by the nonlinear inductor components. The capability of these simulations could be enhanced by including resistors to model the dissipation of the superconducting film, and improving the modelling of discontinuities in transmission line width to take the effects of fringing fields into account. \\

\section*{Data availability statement}
The data that support the findings of this study are available
upon reasonable request from the authors.

\section*{Acknowledgements}
This work was supported by STFC consolidated grant ST/T000414/1. T.Sweetnam is
supported by a STFC PhD studentship, grant ST/T506291/1.

\section*{References}
\bibliographystyle{IEEEtran}
\bibliography{iop_refs}

\end{document}